\documentclass[prb,twocolumn]{revtex4}% Physical Review B

\usepackage{graphicx}% Include figure files
\usepackage{dcolumn}% Align table columns on decimal point
\usepackage{bm}% bold math
\usepackage{color}
\usepackage{multirow}
\usepackage{subfigure}
\usepackage{amssymb}
\usepackage[monochrome]{xcolor}
\begin{document}

\title{Strain effects in monolayer Iron-Chalcogenide superconductors}

\author{Cesare Tresca$^1$, Fabio Ricci$^{2,3}$ and Gianni Profeta$^{1,2}$}
%\affiliation{CNR-SPIN and Dipartimento di Scienze Fisiche e Chimiche, Universit\`a dell'Aquila, 67100 Coppito (L'Aquila), Italy}
\address{$^1$ Dipartimento di Scienze Fisiche e Chimiche, Universit\`a dell'Aquila, 67100 Coppito (L'Aquila), Italy}
\address{$^2$ Consiglio Nazionale delle Ricerche, Istituto Superconduttori, Materiali Innovativi e Dispositivi (CNR-SPIN), 67100 L'Aquila, Italy}
\address{$^3$Physique Th\'eorique des Mat\'eriaux, Universit\'e de Li\`ege (B5), B-4000 Li\`ege, Belgique}

\begin{abstract}
The successful fabrication of one monolayer FeSe on SrTiO$_{3}$ represented a real breakthrough in searching for high-T$_{c}$ Fe-based superconductors~(Ref.~\cite{fese-exp}).
Motivated by this important discovery, we studied the effects of tensile strain on one monolayer and bulk iron-chalcogenide superconductors (FeSe and FeTe), showing that it produces important magnetic and electronic changes in the systems.
We found that the magnetic ground state of bulk and monolayer FeSe is the block-checkerboard phase, which turns into the collinear stripe phase under in plane tensile strain. 
FeTe, in both bulk and monolayer phases, shows two magnetic transitions upon increasing the tensile strain: from  bicollinear in the ground state to  block-checkerboard ending up to the collinear antiferromagnetic phase which could bring it in the superconducting state. 
Finally, the study of the mechanical properties of both FeSe and FeTe monolayers reveals their enormous tensile strain limits and opens the possibility to grow them on different substrates.
\end{abstract}

%\keywords{}

\maketitle

\section*{Introduction}
Since the discovery of high critical temperature superconductivity in fluorine doped iron arsenide LaFeAsO (T$_{c}=$26~K)~\cite{doi:10.1021/ja800073m}, the attention of the scientific community has been captured by Fe-based superconductors (FeSCs). 

\begin{figure}[h]
\includegraphics[width=0.4\linewidth]{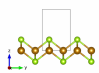}
\includegraphics[width=0.4\linewidth]{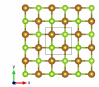}
\hspace{0.35cm}
\includegraphics[width=0.4\linewidth]{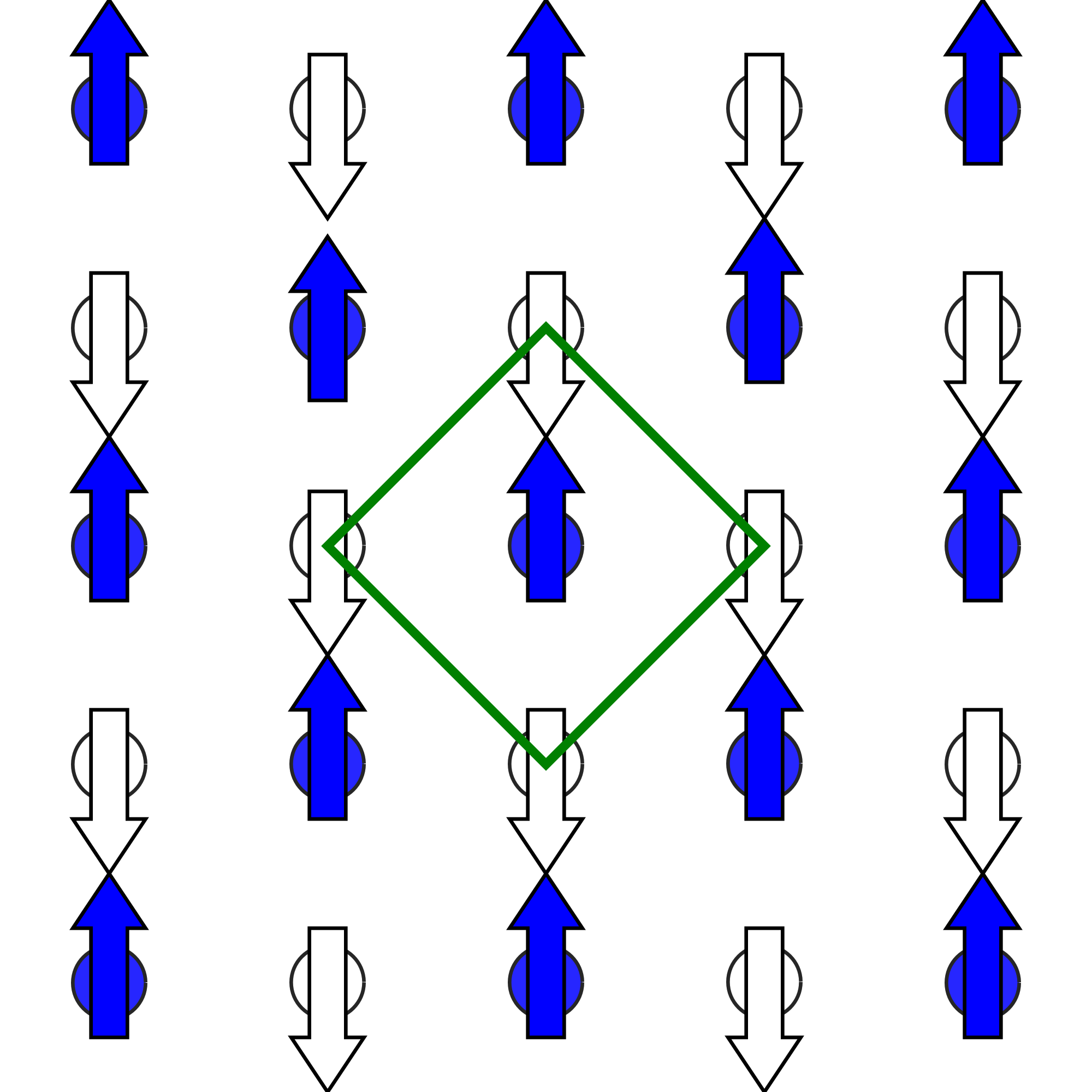}
\hspace{0.35cm}
\includegraphics[width=0.4\linewidth]{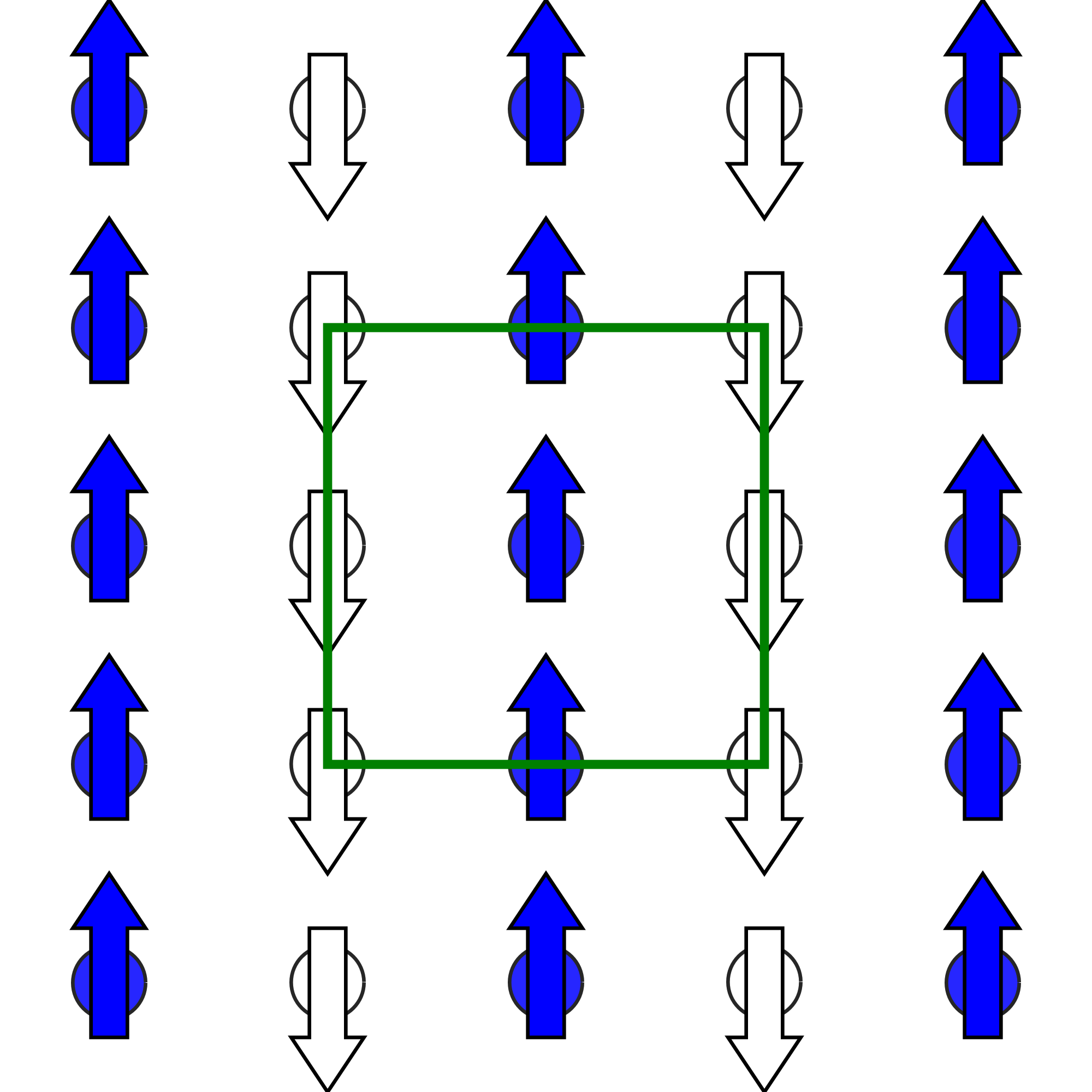}
\hspace{0.35cm}
\includegraphics[width=0.4\linewidth]{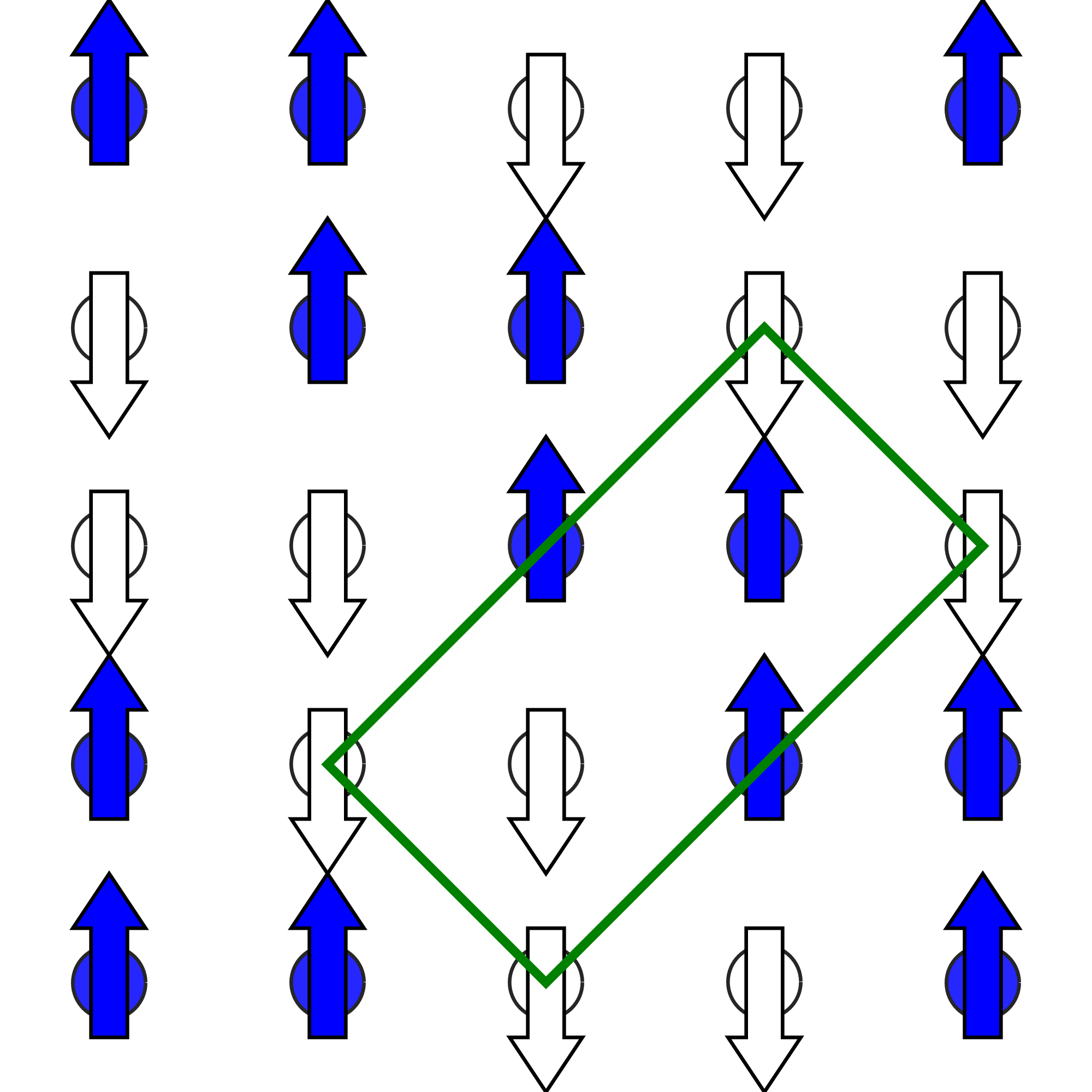}
\hspace{0.35cm}
\includegraphics[width=0.4\linewidth]{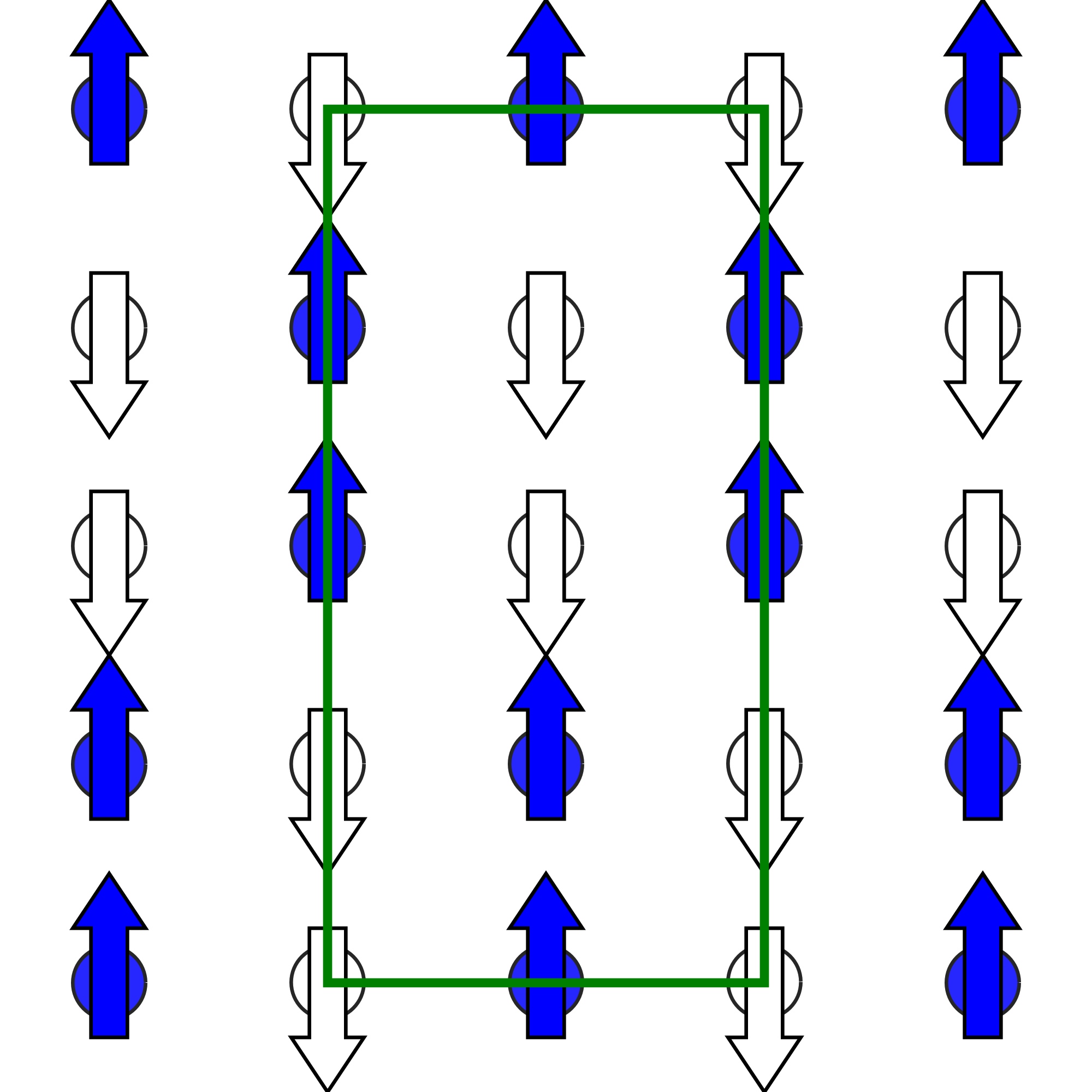}
\hspace{0.35cm}

\caption{Crystal structure of iron chalcogenides and magnetic phases considered. Upper panels: Side (left) and top (right) view of FeCh monolayer crystal. Large spheres indicate Fe atoms while smaller ones Ch atoms.
The unit cell of the bulk phase is drawn as thin line and the directions of in-plane ($\mathbf{a}$ and $\mathbf{b}$) and out-of-plane ($\mathbf{c}$) lattice vectors are highlighted.
Lower panels, from left to right: AFM0, AFM1, AFM2 and AFM3 magnetic phases (see text) with the corresponding in plane unit cell.  }

\label{struct}
\end{figure}

Among all FeSCs families known, the iron-chalcogenide (FeCh) (FeSe, FeTe, FeSe$_{x}$Te$_{1-x}$) family has a fundamental importance for the understanding of the structural, electronic, magnetic and superconducting properties of FeSCs~\cite{Nature_paglione}. In fact, thanks to their simple crystal structure, being composed by one layer of square Fe lattice tetrahedrally bonded to Ch (see Fig.\ref{struct}), FeCh can be thought as a fundamental building block of this class of materials and the perfect test case to study the superconducting mechanisms in these compounds.

FeSe superconducts at ambient conditions with T$_{c}$=8~K, probably due to presence of Se vacancies~\cite{PNAS.105-38}. On the contrary, FeTe is not superconductor, however, partial substitution of Te with Se allows superconductivity to emerge, reaching T$_{c}\sim$15~K at the optimal substitution of 50$\%$. The origin of the superconducting phase is commonly attributed to the developing of spin fluctuations with wave-vector $\mathbf{Q}=(\pi,\pi)$ (associated with a collinear antiferromagnetic stripe phase (AFM1) in which ferromagnetically ordered nearest-neighbouring Fe stripes are aligned perpendicularly to the antiferromagnetically ordered ones, see Fig.\ref{struct})~\cite{Nature_paglione, Wang08042011}. In fact, despite the structural and electronic similarities  with FeSe, FeTe shows a bicollinear antiferromagnetic phase (AFM2)~\cite{Mizuguchi_zkcm_226}.

Structural properties are found to be strongly linked to the superconducting properties in these materials. Indeed, a strong enhancement of T$_{c}$ up to 37~K is observed in FeSe at pressure near 9 GPa~\cite{NatMat8.630}. On the other hand, FeSe thin films under tensile strain of about $3.7\%$ show the suppression of superconducting phase~\cite{Appl.phys.L.94.24}. The  role of the pressure and strain has been discovered to be of fundamental importance also in FeTe. Not only  pressure does not induce superconductivity in FeTe, but favours  a ferromagnetic phase~\cite{PhysRevB.87.094516}. However, it was shown that FeTe thin films become superconductor at 13~K under tensile strain conditions~\cite{PhysRevLett.104.017003}.

A turning point in the exploration of FeSCs was recently provided by the discovery that low dimensionality raises dramatically the superconducting critical temperature of FeSe\cite{fese-exp}. 
Recent experiments, in fact, show exciting properties of monolayer (ML) FeSe, with respect to the bulk phase. In particular,  Meissner effect was observed at 20~K in few layer FeSe grown on SrTiO$_{3}$ (STO) substrate (lattice constant 3.90~\AA) ~\cite{2013arXiv1311.6459D}. Peng \emph{et al.}~\cite{PhysRevLett.112.107001} reported an enhancement of the superconducting critical temperature in the extremely  expanded FeSe monolayer grown on Nb doped STO (lattice constant  3.99~\AA) measuring a superconducting gap up to 70~K. There is direct observation of zero resistivity under 40~K in \emph{ex-situ} experiments~\cite{2014ChPhL..31a7401Z} and $\sim109$~K in the recent \emph{in-situ} experiment with 4-point probe~\cite{2014arXiv1406.3435G}. \\
All these results represent rather solid proofs that a superconducting phase exists indeed in FeSe ML.

The Fermi surface (FS) of monolayer FeSe, measured using \emph{in situ} ARPES technique\cite{Liu_nat, He_nat} also as a function of the film thickness grown on STO\cite{NatMat12.634}, reveals 
that it is characterized by only electron FS at $M$ point of the Brillouin Zone (BZ), apparently losing the hole FS at $\Gamma$. 
Surprisingly, this is topologically different from the FS of the bilayer (and bulk) system, which shows the typical FeSC's FS composed by hole cylinders at the BZ center and electron features at the BZ corners\cite{NatMat12.634}.

At the moment various mechanisms have been proposed to explain the high superconducting transition temperature of the FeSe ML. In particular, the electron-phonon  coupling with oxygen optical phonons in STO which can induce a small momentum pairing further increasing the spin-fluctuation coupling\cite{lee}. The role of the substrate was also emphasized in Ref.\cite{2014arXiv1407.5657C}; here  the authors suggest that the electron-phonon coupling in FeSe is enhanced  by the presence of the substrate imposing  both structural and magnetic constrains.

Indeed, in these particular two dimensional (2D) conditions, an additional role can be played by the substrate and thus by epitaxial strain which may, also, suppress the magneto-elastic coupling.

Considered the reported sensitivity of the physical and superconducting properties of FeCh to structural conditions~\cite{PhysRevLett.104.017003,Appl.phys.L.94.24}, in this paper we studied by first principles density functional theory (DFT)   the effect of strain on the structural, magnetic and electronic properties of FeSe and FeTe in both ML and bulk phases with the aim of understanding the origin of the enhancement  (or of the suppression) of the superconducting phase.

%%%%%%%%%%%%%%%%%%%%%%%%%%%%%%%%%%%%%%%%%%     COMPUTATIONAL DETAILS     %%%%%%%%%%%%%%%%%%%%%%%%%%%%%%%%%%%%%%%
\section{Computational details}\label{comp_dett}
Calculations were performed using the Vienna Ab-initio Simulation Package (VASP)~\cite{PhysRevB.54.11169,Kresse199615}, using the Generalized Gradient Approximation (GGA)~\cite{PhysRevB.59.1758} to the exchange-correlation energy. Van der Waals interactions, not negligible in FeCh systems~\cite{PhysRevB.87.184105},  were included as dispersive term, using the DFT-D2 Grimme's semi-empirical correction~\cite{JCC:JCC20495, vdW}. 

We studied the strain effects considering different magnetic phases (see below). In order to simulate the structural effects of the STO substrate  (which constrains the in-plane lattice constants), we neglected the in-plane magneto-elastic distortion. Once the ground state geometry for each system was determined including magnetic distortions  with inequivalent $\mathbf{a}$ and $\mathbf{b}$ directions,  the systems were constrained to in-plane lattice constants $a=(|{\bf a}| + |{\bf b}|)/2$: this is assumed to be the reference system in the $un-strained$ condition.  Strain effects were studied calculating the total energy as a function of the applied in-plane strain (varying the $a$ in-plane lattice constant) considering different competing magnetic phases (schematically shown in Fig.\ref{struct}): the checkerboard order of the iron spin (called AFM0), the collinear stripe phase (AFM1), the bi-collinear one (AFM2) and the recently proposed block-checkerboard order (AFM3)\cite{afm3}. The magnetic order along the out-of-plane axis is considered ferromagnetic for all the bulk phases.
We used Projected Augmented-Wave  pseudopotentials~\cite{PhysRevB.50.17953} for all the atomic species involved, with an energy cutoff of  350~eV. Integration over BZ was performed using different uniform Monkhorst and Pack grids~\cite{PhysRevB.13.5188} depending on the lattice: $16\times16\times10$ for the tetragonal ($a\times a\times c$) non magnetic (NM) phase (containing two Fe and two Ch atoms), $8\times8\times10$ for magnetic cell  ($2a\times2a\times c$) commensurate to the AFM0, AFM1 and AFM2 magnetic phases. The AFM3 phase was simulated using $\sqrt2a\times2\sqrt2a\times c$ unit cell using a Monkhorst and Pack grid of  $8\times4\times10$.
For the ML systems we considered a supercell including $\sim10$~\AA~of vacuum. Using these parameters the total energy of the considered systems are converged to less than 0.1 meV/atom.
%A Gaussian smearing with $\sigma=0.02$  eV was considered for all the calculations. 

The calculations of  band structures and FS's have been performed using the NM electronic states\cite{PhysRevB.83.094529}.

%%%%%%%%%%%%%%%%%%%%%%%%%%%%%%%%%%%%%%%%%%     RESULTS AND DISCUSSION     %%%%%%%%%%%%%%%%%%%%%%%%%%%%%%%%%%%%%%%
\section{Results and discussion}\label{result}

\begin{figure}
\includegraphics[width=0.8\linewidth]{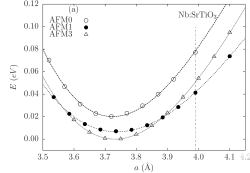}\label{Se_ML_mag}
\includegraphics[width=0.8\linewidth]{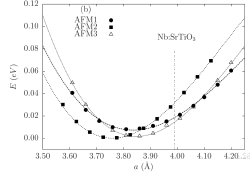}\label{Te_ML_mag}% 
\caption{Total energy per atom as a function of the in-plane lattice parameter for ML FeSe (panel (a)) and FeTe (panel (b)). The theoretical points are fitted with a cubic function. 
The dotted vertical line indicates the Nb:STO lattice parameter (see text).}\label{Te_Se_ML_mag}
\end{figure}

The total energy of FeSe ML as a function of the in-plane lattice constant for the lowest energy magnetic structures are shown in Fig.~\ref{Te_Se_ML_mag}(a). At variance with what was expected in line with the collinear stripe ground state of pnictides, the  ground state magnetic order is the block-checkerboard with equilibrium structural parameters  reported in Tab.~\ref{tab1}.
Similar with what found in the AFM1 phase, the magneto-elastic coupling causes the inequivalence of $\mathbf{a}$ and $\mathbf{b}$ directions. 

The magnetic moment of Fe atoms in the monolayer phase is 0.2 $\mu_B$ higher with respect to the bulk system (see Table \ref{tab1}) and changes slightly (less than 0.1 $\mu_B$) depending on the magnetic order.

Fig.\ref{Te_Se_ML_mag}(a) shows that the AFM1 phase is the more stable phase under  applied  in-plane strain, which induces a magnetic phase transition at $a\simeq 3.9$~\AA.
Considering that the superconducting pairing could be mediated by $(\pi,\pi)$ spin-fluctuations, free-standing stoichiometric FeSe monolayer with a $(\pi,\pi/2)$ magnetic order is not expected to be superconductor, however it may support a superconducting phase if strained at the STO lattice constant in agreement with  experimental evidences\cite{PhysRevLett.112.107001}.
The AFM1 phase increases its stability under applied strain and becomes the ground state at the Nb:STO lattice constant  (see Fig.\ref{Te_Se_ML_mag}(a)) where experiments report the highest critical temperature\cite{PhysRevLett.112.107001}.

\begin{table}[h]
\begin{center}
\begin{tabular}{|l|c|c|}
\hline
		& FeSe    & FeTe \\
\hline
$|{\bf a}|~$(\AA)	& 3.75   & 4.04	 \\

$|{\bf b}|~$(\AA)	& 3.72	   & 3.51	\\

$h_{ch}~$(\AA)	& 1.44	   & 1.79	\\

%$\alpha_{leg}~(deg)$& 104.8	   & 93.0	\\

$|m|$ $(\mu_{B})$ & 2.1 & 2.5  \\

\hline
\end{tabular}
\end{center}
\caption{Equilibrium structural parameters of FeSe and FeTe ML in their respective ground state ($|{\bf a}|$ and $|{\bf b}|$ lattice constants and the height of the chalcogenide, $h_{ch}$) and Fe magnetic moment, $|m|$.}
\label{tab1}
\end{table}

\begin{figure}[h]
\begin{center}
\subfigure[ $~a=3.75$~\AA.\label{FS_Se_ML_0}]%
{\includegraphics[clip,width=0.4\linewidth]{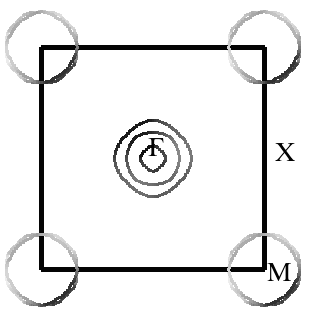}}
\subfigure[ $~a=3.99$~\AA.\label{FS_Se_ML_99}]%
{\includegraphics[clip,width=0.4\linewidth]{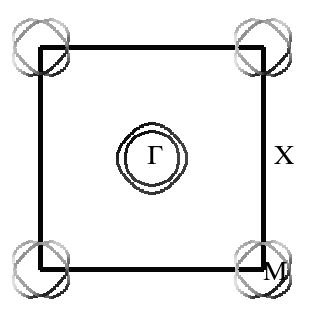}}% 
\caption{Fermi surface of FeSe at equilibrium (a) and strained at Nb:STO lattice parameter (b). }
\label{FS_Se_ML}
\end{center}
\end{figure}

The variation of about $6\%$ of the in-plane lattice constant (Nb:STO) determines a strong effect on the FS, as reported in Fig.~\ref{FS_Se_ML}.  This is mainly due to the contraction of the Se height from the Fe plane, h$_{ch}$ (Fig.\ref{struct}), which moves from 1.44~\AA~(equilibrium geometry) to 1.34~\AA~ at the in-plane lattice constant of Nb:STO. 
In the free-standing ML, there are three hole-type FS's at the $\Gamma$ point and two (quasi overlapping) electron-like FS's at the BZ corner (Fig.~\ref{FS_Se_ML_0}). Stretching the FeSe ML at the Nb:STO lattice parameter (Fig.~\ref{FS_Se_ML_99}) fills completely the $d_{x^2-y^2}$ hole-type band near $\Gamma$ and makes the 
corresponding FS  to vanish. At the same time, the electron FS's change their shape and shrink as a result of electron counting. ARPES measurements, however, seem to find a different scenario~\cite{NatMat12.634}: only electron sheets at the BZ corner are found with no evidences of  hole-type FS at the BZ center.

Although the calculated electronic properties do not completely agree with available experiments, we indeed predict that strain induces a magnetic phase transition and a subsequent topological transition in the FS of the ML phase of FeSe.

We now consider the monolayer phase of FeTe, a system not yet experimentally realized.  
As already discussed, bulk FeTe is not superconducting, however the effect of the in-plane strain has been shown to induce a superconducting phase with a critical temperature of 13~K~\cite{PhysRevLett.104.017003}. In principle, a monolayer of FeTe can result more interesting than the FeSe monolayer; in fact, if superconductivity is mediated by spin-fluctuations, FeTe is expected to have a stronger pairing~\cite{PhysRevB.78.134514}.

The ground state of FeTe ML is found to be AFM2 with lattice parameters reported in Tab.~\ref{tab1}.  While  the magnetic induced distortion of the square lattice is 
minimal in FeSe, in FeTe it is sensible. In addition,  
the magnetic moment of Fe atoms is larger in the ground state (AFM2) of FeTe  than in FeSe (see Table \ref{tab1}) and shows a reduction of 0.3$\mu_B$ in the AFM1 phase.
From Fig.~\ref{Te_Se_ML_mag}(b) we note that two magnetic phase transitions take place as a function of the tensile strain: from AFM2 to AFM3 at  $a\gtrsim3.80$~\AA\ and from AFM3 to AFM1 at  $a\gtrsim4.05$~\AA.

\begin{figure}[h]
\begin{center}
\subfigure[ $~a=3.78$~\AA.\label{FS_Te_ML_0}]%
{\includegraphics[width=0.4\linewidth]{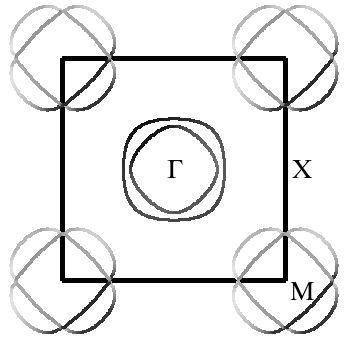}}
\subfigure[ $~a=4.17$~\AA.\label{FS_Te_ML_417}]%
{\includegraphics[width=0.395\linewidth]{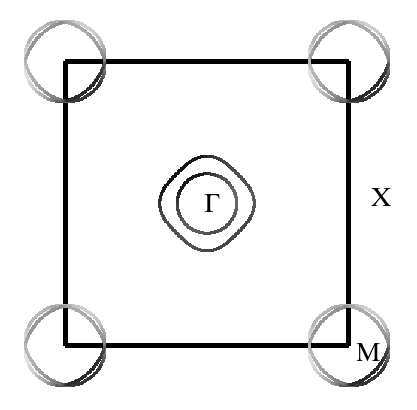}}% 
\caption{Fermi surface for FeTe ML at equilibrium (panel a) and strained at 4.17~\AA~ in-plane lattice parameter (panel b). }\label{FS_Te_ML}
\end{center}
\end{figure}

The FS's shape of FeTe ML changes from the equilibrium (Fig.\ref{FS_Te_ML_0}) to the strained phase (Fig.~\ref{FS_Te_ML_417}). In this last geometry, the FS  shows  
two hole pockets at $\Gamma$ and two (quasi overlapping) electron features at $M$ nearly nested with $\mathbf{Q}=(\pi,\pi)$.
Therefore the magnetic transition towards the AFM1 phase with this ordering vector allows the pairing between  hole and electron FS's sheets which was hindered in the
  bicollinear and block-checkerboard phases.
Based on these structural, magnetic and electronic properties we propose   strained  FeTe ML as  a  good candidate for interesting superconducting properties, potentially better than FeSe ML.

The results obtained in the monolayer systems help the interpretation of the effects of in-plane strain on the superconducting properties of FeCh bulk systems.
Indeed, for FeSe and FeTe thin films a different behaviour under tensile strain was experimentally reported. FeSe grown on STO shows suppression of the superconducting phase observed on unstrained compounds\cite{Appl.phys.L.94.24}, while, bulk FeTe becomes superconductor upon the same tensile strain conditions\cite{PhysRevLett.104.017003}.

\begin{figure}[h]
\begin{center}
\subfigure[$~a=3.73$~\AA.\label{FS_Se_B_0}]%
{\includegraphics[width=6cm]{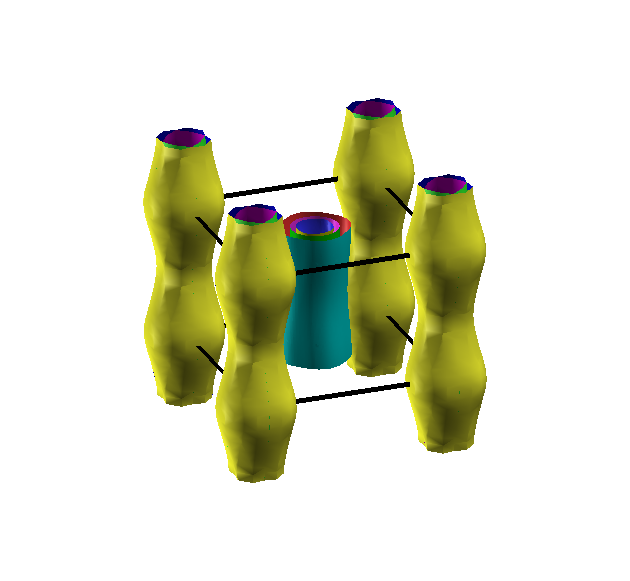}}
\subfigure[$~a=3.99$~\AA.\label{FS_Se_B_99}]%
{\includegraphics[width=6cm]{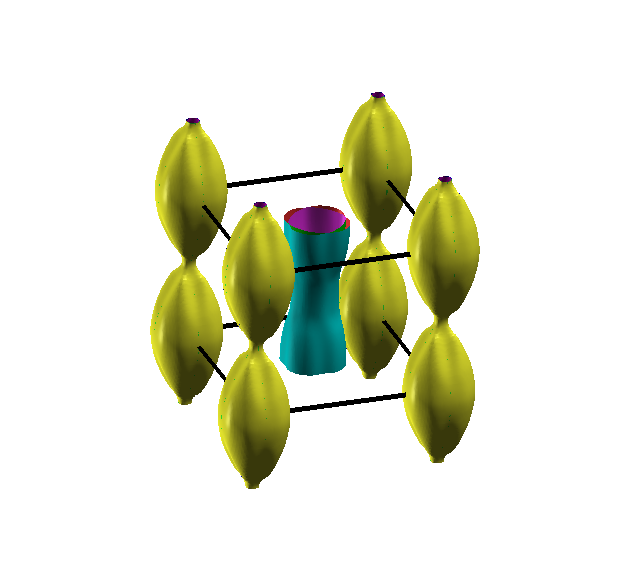}}% 
\caption{Fermi surface for FeSe bulk at equilibrium (panel (a)) and strained at Nb:STO lattice parameter (panel (b)). The $\Gamma$-point is the center of the first BZ shown as thin lines.}\label{FS_Se_B}
\end{center}
\end{figure}

Total energy calculations performed constraining the in-plane lattice parameter and fully relaxing all the others ($i.e.$ out-of-plane axis  and internal positions) show 
that the magnetic ground state of bulk FeSe at equilibrium is the AFM3 order, and it becomes AFM1 at the Nb:STO in-plane lattice constant.
Thus, suppression of the superconducting state in strained FeSe~\cite{Appl.phys.L.94.24,winiarski_fese_FS} can be interpreted 
considering that strain causes a topological transition leading to a complete closure of the electron FS cylinders  
at $M$ (present at the equilibrium, see Fig.\ref{FS_Se_B_0}). At $a\gtrsim3.99$~\AA~ the FS becomes three dimensional 
 (see Fig.~\ref{FS_Se_B_99}),  strongly reducing the pairing. 

\begin{figure}[h]
\begin{center}
\subfigure[ $~a=3.75$~\AA.\label{FS_Te_B_0}]%
{\includegraphics[width=6cm]{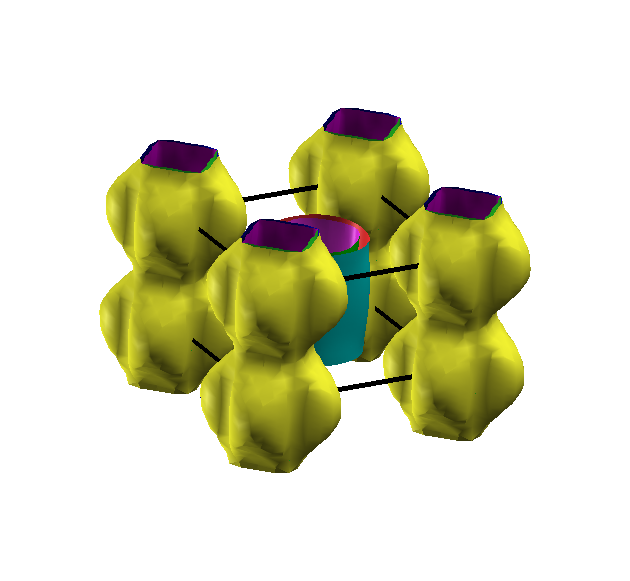}}
\subfigure[ $~a=4.10$~\AA.\label{FS_Te_B_410}]%
{\includegraphics[width=6cm]{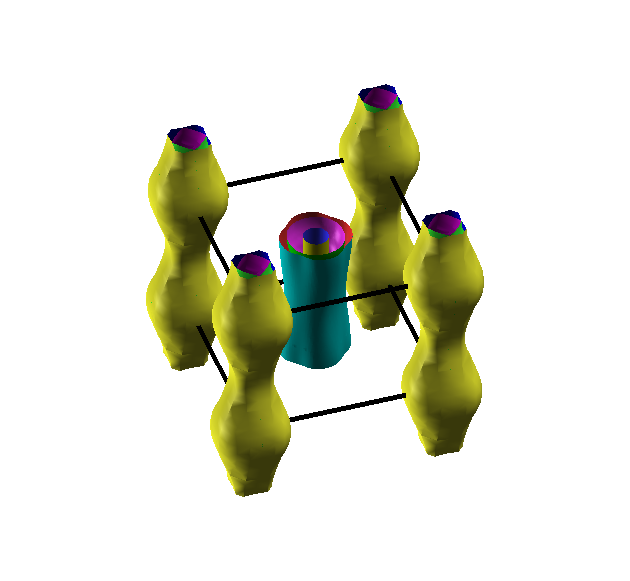}}% 
\caption{Fermi surface for FeTe bulk  at equilibrium (panel (a)) and strained at 4.10~\AA~ lattice parameter (panel (b)). 
The $\Gamma$-point is the center of the first BZ shown as thin lines.}\label{FS_Te_B}
\end{center}
\end{figure}

On the contrary, we find that FeTe bulk has a bicollinear (AFM2) magnetic ground state in agreement with experiments\cite{PhysRevLett.102.247001}. However, a transition towards the AFM3 phase takes place at $a\simeq3.8$~\AA; upon further increase of  the tensile strain above $a\gtrsim4.10$~\AA, the system develops an AFM1 ground state. In this last phase, bulk FeTe preserves both hole and electron sheets although with sensible differences with respect to the equilibrium phase (Fig.\ref{FS_Te_B}).  
On the basis of these results, the experimental evidences reported by Han {\it et al.}~\cite{PhysRevLett.104.017003} find now a natural interpretation:  strain causes AFM1 spin-fluctuations which couple the hole and electron FS sheets via the AFM1 {\bf Q} vector.

\begin{figure}[h]
\begin{center}
\includegraphics[width=0.7\linewidth]{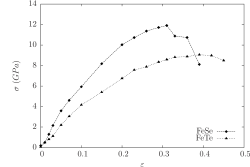}
\caption{Stress-strain relation for FeSe and FeTe ML.}\label{stress_strain_Se_Te}
\end{center}
\end{figure}

In order to possibly guide new experiments towards the use of alternative substrates to properly tune strain conditions in both bulk and monolayer systems,
 we investigated the mechanical properties of FeSe and FeTe ML to obtain information on the stiffness and breaking strength of these compounds.
Starting from the equilibrium structure, we studied the stress-strain relation for FeSe and FeTe ML and report the results in Fig.~\ref{stress_strain_Se_Te}. 
Here, the  tensile strain  is defined as $\varepsilon=(\varepsilon_{xx}+\varepsilon_{yy})/\sqrt{2}$ with $\varepsilon_{xx}=(a'-a)/a$ and $\varepsilon_{yy}=(b'-b)/b$~\cite{PhysRevB.89.184111}, being $a$ and $b$  the lattice parameters along $\hat{x}$ and $\hat{y}$ directions respectively~\cite{redef_stress}.

The stress-strain curve for  FeSe and FeTe  shows a maximum at the critical strain of $\varepsilon=0.33$ and  $\varepsilon=0.39$, respectively. Although the corresponding maximal stress ($\sigma^{FeSe}_{max}=11.9$~GPa and $\sigma^{FeTe}_{max}=9.1$~GPa) is very low compared to other monolayers as graphene or  dichalcogenides (\emph{i.e.} MoS$_{2}$), these results demonstrate that both compounds can support relevant strain, comparable with what  recently found for phosphorene~\cite{2014arXiv1403.7882W}. This  can open the way to future growth of superconducting chalcogenides on different substrates with quite 
large mismatch.

\section{Conclusions}\label{conclusion}

Using first-principles DFT simulations we calculated the mechanical, electronic and magnetic properties of FeCh in 2D and bulk geometries under in-plane strain.
Our predictions show that: 

$(i)$ The magnetic ground state of monolayer and bulk FeSe is the block-checkerboard. Application of tensile strain produces a magnetic transition towards the AFM1 phase with the closure of only one FS at the $\Gamma$ point in FeSe ML. 

In the bulk geometry, strain induces a topological transition of the FS  at $M$, that can be at the origin of the suppression of the superconducting state. 

$(ii)$ If FeTe ML could be strained above $9\%$, it will achieve all the requirements needed  to observe a superconducting transition:  AFM1 magnetic ground state with hole and electron FS's sheets nested with the proper ordering wave vector. 
The same effect is predicted also in the bulk phase as well.

$(iii)$ Both FeSe and FeTe ML have large enough mechanical flexibility to support critical strains up to $\varepsilon\gtrsim 30\%$.
\section{Acknowledgments}
We thank Prof. L. Ottaviano for stimulating discussions and for valuable editing contribution and Prof. A. Continenza for critically reading the manuscript.
This work was supported by FP7 European project SUPER-IRON (grant agreement No. 283204) and by the Italian Ministry of University Research through the PRIN 2012 project.
We also acknowledge the CINECA award under the ISCRA initiative, for access to high performance computing resources and support.

\vspace{10pt}
\bibliography{bibliography}

\begin{thebibliography}{10}
\bibitem{vdW}
  We considered the equilibrium structure obtained considering the vdw
  correction only for the out-of-plane lattice constants (see also Ref.
  \cite{PhysRevB.87.184105}).
\bibitem{redef_stress}
  In order to calculate the stress in 2D, we divided the computed 3D stress by
  a factor $c/d_0$, where $c$ is the lattice vector of the supercell used in
  simulation and $d_0$ is the distance between the Fe planes in the
  corresponding bulk system.
\end{thebibliography}

\end{document}